# Derivation of the Schrödinger Equation and the Klein-Gordon Equation from First Principles

Gerhard Grössing
*Austrian Institute for Nonlinear Studies*
Parkgasse 9, A-1030 Vienna, Austria

**Abstract**: The Schrödinger- and Klein-Gordon equations are directly derived from classical Lagrangians. The only inputs are given by the quantization of energy ($E = \hbar\omega$) and momentum ($p = \hbar k$), respectively, as well as the assumed existence of a space-pervading field of "zero-point energy" ($E_0 = \hbar\omega/2$ per spatial dimension) associated to each particle of energy $E$. The latter leads to an additional kinetic energy term in the classical Lagrangian, which alone suffices to pass from classical to quantum mechanics.
Moreover, Heisenberg's uncertainty relations are also derived within this framework, i.e., without referring to quantum mechanical or other complex-numbered functions.

## 1. Introduction

The fundamental equations of quantum theory, like the Schrödinger equation or its relativistic analogues, are usually put forward on heuristic grounds only, i.e., they are not derived from an underlying canonical set of axioms. Schrödinger himself arrived at the equation named after him by simply inserting de Broglie's relation (i.e., between the momentum of a particle and its associated wavelength) into a classical wave equation. [1] The only attempt to strictly *derive* the Schrödinger equation from classical physics, for example, on the basis of a new differential calculus, is due to Nelson's stochastic theory [2,3]. However, apart from being a purely local theory, which is thus at odds with the phenomenon of quantum mechanical nonlocality, Nelson's calculus itself, like, e.g., the mathematical arbitrariness of the definition of acceleration, has not been accepted as a reliable foundational basis of quantum theory.

Recently, Hall and Reginatto [4] showed that the Schrödinger equation can be derived from an "exact uncertainty principle" by starting off with assumed momentum fluctuations of a particle. The latter will also be important in our derivation, but we shall try here to arrive at the Schrödinger equation without assuming some uncertainty principle, which actually already is a genuine part of quantum physics. In contrast, once the Schrödinger equation will be shown here to result from a particular modification of classical physics, also the uncertainty principle can be shown to follow in a straightforward manner.

In fact, in this paper the Schrödinger- and Klein-Gordon equations are directly derived from classical Lagrangians. The only inputs will be given by the empirical evidence for quantized expressions for energy ($E = \hbar\omega$) and momentum ($p = \hbar k$), respectively, as well as the existence of an associated space-pervading field with an average "zero-point energy" of $E_0 = \hbar\omega/2$ per spatial dimension.[5] In other words, we shall assume in the following that the energy of the total system of "particle plus ($n-$dimensional) environment" is given by the time derivative of the action function $S$,

$$\frac{\partial S}{\partial t} = -\left(\hbar\omega + n\frac{\hbar\omega}{2}\right). \qquad (1.1)$$

Thus, next to the usual action function describing a (classical) physical system (including here one particle of energy $E = \hbar\omega$), the existence of an "environment" of the particle in the form of a zero-point background field implies a separate term in the action function for any point $\mathbf{x}$ in $n-$dimensional space, i.e.,

$$S_0(\mathbf{x},t) := -n\int \frac{\hbar\omega(\mathbf{x},t)}{2}dt. \qquad (1.2)$$

It is thus assumed that the zero-point energy pervades all of a particle's surrounding space in the form of a field of harmonic oscillators. [6] Upon higher spatial resolution than is usually needed in classical physics, said field may also be imagined as a "rugged landscape" of a multitude of oscillating elements. This will lead to energy gradients and additional momenta $\nabla S_0$ on said "microscopic" scale. One can reasonably assume that the additional dynamics will usually be governed by a principle of least energy, or, more exactly, an extremal principle, at any time. However, it will turn out that for our purposes an assumed time-integrated extremal principle will suffice, i.e.,

$$\int \delta\left(\frac{\partial S_0}{\partial t}\right)dt = 0. \qquad (1.3)$$

This is also equivalent to the assumption that eventual short-term variations $\delta\omega = \sum_i \frac{\partial \omega}{\partial x_i}\delta x_i$ on the average will cancel each other out when integrated over sufficiently large periods of time. Thus, only the integrated expressions of equation (1.2) may lead to non-vanishing gradients $\nabla S_0$. (In fact, one could also consider more restrictive alternatives to (1.3), which thus hints at a broader basis of the approach presented here for possible modellings of quantum systems and which may be a fruitful topic for future research.)

Given this extra zero-point-field, also an extra term for the kinetic energy of a particle is implied. In fact, this will be the only necessary assumption next to equation (1.2) to derive the Schrödinger equation from a variation principle involving a "classical" Lagrangian, as shall be shown now.

## 2. Derivation of the Schrödinger Equation

In the Hamilton-Jacobi formulation of classical mechanics, the action integral for a single particle in an $n$-dimensional configuration space, with some external potential $V$, is given by [7]

$$A = \int L_C d^n x\, dt = \int P \left\{ \frac{\partial S}{\partial t} + \frac{1}{2m} \nabla S \cdot \nabla S + V \right\} d^n x\, dt, \qquad (2.1)$$

where the action function $S(\mathbf{x},t)$ is related to the particle velocity $\mathbf{v}(\mathbf{x},t)$ via

$$\mathbf{v} = \frac{1}{m} \nabla S, \qquad (2.2)$$

and the probability density $P(\mathbf{x},t)$ that a particle is found in a given volume of configuration space is normalized such that

$$\int P d^n x = 1. \qquad (2.3)$$

Upon fixed end-point variation, i.e., $\delta P = \delta S = 0$ at the boundaries, of the Lagrangian in equation (2.1), one obtains the continuity equation for the probability density

$$\frac{\partial P}{\partial t} + \nabla \cdot \left( P \frac{\nabla S}{m} \right) = 0 \qquad (2.4)$$

and the equation of motion ("Hamilton-Jacobi equation")

$$\frac{\partial S}{\partial t} + \frac{1}{2m} \nabla S \cdot \nabla S + V = 0. \qquad (2.5)$$

Now we propose the following extension of the classical theory. Due to the assumed activity of the zero-point background field producing an additional, non-classical dynamics, $\mathbf{p} = \nabla S$ is only an average momentum which is subject to momentum fluctuations $\mathbf{f}$ around $\nabla S$. Thus, the physical momentum is given by

$$\mathbf{p} := \nabla S + \mathbf{f}. \qquad (2.6)$$

An equation of the form (2.6) is also the starting point in the derivation of Hall and Reginatto [4]. However, these authors do not assume a particular physical model for the momentum fluctuations. In fact, they tend to "regard the fluctuations as fundamentally nonanalyzable, being introduced as a simple device to remove the notion of individual particle trajectories". Still, as shall be shown here, one may very well retain the notion of particle trajectories and consider a physically motivated ansatz of the form (2.6) to show that it is exactly this which can lead from classical to quantum physics.

We proceed by a further agreement with Hall and Reginatto in that the momentum fluctuations $\mathbf{f}$ are assumed to be linearly uncorrelated with the average momentum $\mathbf{p} = \nabla S$. In other words, with the fluctuations themselves being considered unbiased, i.e., $\int P\mathbf{f}d^n x = 0$, the average over fluctuations *and* position of the product of the two momentum components are also assumed to vanish identically [4]:

$$\int P(\nabla S \cdot \mathbf{f})d^n x = 0. \tag{2.7}$$

Thus, it is proposed that the complete action integral of a particle immersed in the zero-point field is given by

$$A := \int P\left\{\frac{\partial S}{\partial t} + \frac{\mathbf{p} \cdot \mathbf{p}}{2m} + V\right\}d^n x dt, \tag{2.8}$$

with $\mathbf{p}$ given by equation (2.6). With condition (2.7), this provides

$$A = \int P\left\{\frac{\partial S}{\partial t} + \frac{1}{2m}\nabla S \cdot \nabla S + V\right\}d^n x dt + \frac{1}{2m}\int (\Delta f)^2 dt, \tag{2.9}$$

where $\Delta f$ is the average rms momentum fluctuation, i.e.,

$$(\Delta f)^2 = \int P\mathbf{f} \cdot \mathbf{f} d^n x. \tag{2.10}$$

Now we introduce our assumption stated above, i.e, that the additional momentum fluctuation $\mathbf{f}$ is due to the activity of the zero-point energy field described by the additional momentum generating function $S_0$ of equation (1.2) such that

$$(\Delta f)^2 = \int P(\nabla S_0)^2 d^n x. \tag{2.11}$$

Thus, combining equations (2.9) and (2.11), we obtain the complete action integral of a particle immersed in the zero-point field as

$$A := \int P\left\{\frac{\partial S}{\partial t} + \frac{(\nabla S)^2}{2m} + \frac{(\nabla S_0)^2}{2m} + V\right\}d^n x dt. \tag{2.12}$$

Now we perform the fixed end-point variation in $S$ of the Euler-Lagrange equation, i.e.,

$$\frac{\partial L_C}{\partial S} - \frac{\partial}{\partial x_i}\left\{\frac{\partial L_C}{\partial\left(\frac{\partial S}{\partial x_i}\right)}\right\} = 0, \qquad (2.13)$$

where the index $i$ runs over the time and the three spatial components, respectively. Elaborating (2.13), we obtain the usual continuity equation: With the definition of the *material derivative*, i.e., the time rate of change of a function while moving with the particle, as

$$\frac{d}{dt} := \frac{\partial}{\partial t} + \mathbf{v}\cdot\nabla, \qquad (2.14)$$

one obtains along a path that

$$\frac{dP}{dt} + (\nabla\cdot\mathbf{v})P = 0, \qquad (2.15)$$

with the solutions

$$P(\mathbf{x},t) = P(\mathbf{x}_0,t_0)\exp\left\{-\int(\nabla\cdot\mathbf{v})_\mathbf{x}\,dt\right\}. \qquad (2.16)$$

As to the integral in (2.16), we note that the only movement of the particle deviating from the classical path must be due to the zero-point background as given in equations (1.2) and (2.11), respectively. With the motion of a harmonic oscillator being defined by its frequency $\omega$, the corresponding additional undulatory movement provides for any one such oscillator at some location $\mathbf{x}$ that for $n=3$ dimensions

$$\nabla\cdot\mathbf{v} = \frac{\partial}{\partial x}(\omega x) + \frac{\partial}{\partial y}(\omega y) + \frac{\partial}{\partial z}(\omega z) = 3(\omega + l\delta\omega), \qquad (2.17)$$

where, as noted in the introduction, the second term is assumed to vanish, i.e., the variations $\delta\omega = \frac{\partial\omega}{\partial x_i}\delta x_i$, with $x_i = x, y, z$, with equal weights $l = \frac{x_i}{\delta x_i}$, for all $i$, on the average cancel each other out due the extremal principle of equation (1.3).

Thus, one obtains with equations (2.16) and (1.2) that in general

$$P(\mathbf{x},t) = P(\mathbf{x}_0,t_0)\exp\left\{-n\int\omega(\mathbf{x},t)dt\right\} = P(\mathbf{x}_0,t_0)\exp\left\{2S_0(\mathbf{x},t)/\hbar\right\}. \qquad (2.18)$$

Finally, from equation (2.18) we derive the expression for our new velocity fluctuation field $\mathbf{u} := \frac{1}{m}\nabla S_0$ as

$$\mathbf{u} = \frac{\hbar}{2m}\frac{\nabla P}{P}. \qquad (2.19)$$

Moreover, fixed end-point variation of equation (2.12) in $S_0$ provides with (2.19) that

$$\frac{\partial P}{\partial t} = -\frac{\hbar}{2m}\nabla^2 P, \qquad (2.20)$$

which is reminiscent of Fick's law for the propagation of diffusion waves. In fact, it can be noted here that the above *derived* expression (2.19) is exactly identical to the *osmotic velocity* of the stochastic interpretation of quantum mechanics. [2,3] Moreover, one can thus illustrate clearly the linear independence of the two terms, $\nabla S$ and $\mathbf{f}$, in equation (2.7), as $\mathbf{f}$ must be associated with an "orthogonal" osmotic process of unbiased momentum fluctuation contributions, rather than with a momentum component simply to be added to $\nabla S$ on the basis of some shared propagation law.

If we now insert equation (2.19) into the action integral (2.12), we obtain as its final form

$$A = \int P\left\{\frac{\partial S}{\partial t} + \frac{(\nabla S)^2}{2m} + \frac{m}{2}\left(\frac{\hbar}{2m}\frac{\nabla P}{P}\right)^2 + V\right\}d^n x\, dt. \qquad (2.21)$$

Note that this is identical with the classical expression (2.1) except for the third term. Performing now the fixed end-point variation in $P$ of the Euler-Lagrange equations

$$\frac{\partial L_C}{\partial P} - \frac{\partial}{\partial x_i}\left\{\frac{\partial L_C}{\partial\left(\frac{\partial P}{\partial x_i}\right)}\right\} = 0, \qquad (2.22)$$

one obtains the so-called Hamilton-Jacobi-Bohm equation [8,9], i.e.,

$$\frac{\partial S}{\partial t} + \frac{(\nabla S)^2}{2m} + V + \frac{\hbar^2}{4m}\left[\frac{1}{2}\left(\frac{\nabla P}{P}\right)^2 - \frac{\nabla^2 P}{P}\right] = 0. \qquad (2.23)$$

However, as is well known [8,9], the equations (2.15) and (2.23), together with the introduction of the complex-numbered "wave function"

$$\psi = \sqrt{P}\exp\{-iS/\hbar\}, \qquad (2.24)$$

can be condensed into a single equation, i.e., the Schrödinger equation

$$i\hbar\frac{\partial \psi}{\partial t} = \left(-\frac{\hbar^2}{2m}\nabla^2 + V\right)\psi. \tag{2.25}$$

Extension to a many-particle system is straightforwardly achieved by starting the same procedure with a correspondingly altered Lagrangian in (2.12), which then ultimately provides the usual many-particle Schrödinger equation.

### 3. Derivation of the Klein-Gordon Equation

We can now proceed to the relativistic (spinless) case. With the four-vector notation $dx^\mu := (cdt, d\mathbf{x})$, with the usual sum convention, and in accordance with relativistic kinematics, the action integral for a free particle in a four-dimensional volume $\Omega$ of phase space can be formulated as

$$A = \int L d^4\Omega = \int P\left\{\frac{1}{m}p_\mu p^\mu - E\right\} d^4\Omega. \tag{3.1}$$

Moreover, introducing in complete analogy to the nonrelativistic case an extra kinetic energy term involving a velocity fluctuation field $u_\mu := \frac{\partial_\mu S_0}{m}$, with $S_0$ again given by the (Lorentz-invariant) expression (1.2), the Lagrangian becomes

$$L = P\left\{\frac{1}{m}\partial_\mu S \partial^\mu S + \frac{1}{m}\partial_\mu S_0 \partial^\mu S_0 + \frac{\partial S}{\partial t}\right\}. \tag{3.2}$$

(Note that the Lagrangian is in many cases covariant only in the absence of an external potential, which is why we here consider the free-particle case only. However, as in the nonrelativistic situation, extension to the many-particle case is straightforward.)

Fixed end-point variation in $S$, i.e.,

$$\frac{\partial L}{\partial S} - \partial_\mu \left\{\frac{\partial L}{\partial(\partial_\mu S)}\right\} = 0, \tag{3.3}$$

then provides the covariant continuity equation

$$\partial_\mu\left[P \partial^\mu S\right] = 0. \tag{3.4}$$

Since equations (2.15) and (2.18), respectively, also follow from equation (3.4), we obtain from equation (2.18) that

$$\frac{\partial_\mu P}{P} = \frac{2}{\hbar}\partial_\mu S_0, \qquad (3.5)$$

and thus

$$u_\mu = \frac{\partial_\mu S_0}{m} = \frac{\hbar}{2m}\frac{\partial_\mu P}{P}. \qquad (3.6)$$

Inserting the latter into the Lagrangian (3.2), and performing the variation in $P$, i.e.,

$$\frac{\partial L}{\partial P} - \partial_\mu \left\{\frac{\partial L}{\partial(\partial_\mu P)}\right\} = 0, \qquad (3.7)$$

finally provides for $\frac{\partial S}{\partial t} := -mc^2$:

$$\partial_\mu S \partial^\mu S = m^2 c^2 + m^2 u_\mu u^\mu + m\hbar \partial_\mu u^\mu. \qquad (3.8)$$

As the last two expressions on the r.h.s. of (3.8) are identical to the relativistic expression for the "quantum potential" term [10], we have obtained the relativistic Hamilton-Jacobi-Bohm equation, i.e.,

$$\partial_\mu S \partial^\mu S := M^2 c^2 = m^2 c^2 + \hbar^2 \frac{\Box\sqrt{P}}{\sqrt{P}}. \qquad (3.9)$$

Again, as is well known, equations (3.4) and (3.9) can be written in compact form by using the "wave function" $\Psi = \sqrt{P}\exp(-iS/\hbar)$ to obtain the usual Klein-Gordon equation

$$\left(\Box + \frac{m^2 c^2}{\hbar^2}\right)\Psi = 0. \qquad (3.10)$$

We have thus succeeded in deriving the Schrödinger- and Klein-Gordon equations from classical Lagrangians with a minimal set of additional assumptions relating to the zero-point energy field associated to each particle.

## 4. Conclusions and Outlook

The derivation of fundamental quantum mechanical equations from classical (real-valued) Lagrangians suggests the possibility to obtain all results of quantum theory without ever making use of complex "probability amplitudes" ("wave functions"). This may not always be practical, and is in no way an argument against the well-established machinery of the standard quantum mechanical formalism. However, in the foundational debate, it sheds new light on old problems. For example, one can derive Heisenberg's uncertainty relations without invoking complex wave functions, as shall be shown now (i.e., in one spatial dimension, for simplicity).

We have seen that the introduction of an "osmotic" velocity fluctuation field (2.19) is necessary to obtain a complete description of a "particle immersed in the zero-point field". So, even if for the time being we assumed that our knowledge of the particle's momentum were given in one part by the exact classical momentum (i.e., with infinite precision), we must still consider the latter to be "smeared" by the presence of the "osmotic" momentum fluctuation $\mathbf{f}$ in equation (2.6), such that the uncertainty in the particle's momentum $\delta p_0$ would then be given by the average rms momentum fluctuation according to equation (2.11), which with equation (2.19) becomes

$$\delta p_0 := \Delta f = \sqrt{\int P \left( \frac{\hbar}{2} \frac{\nabla P}{P} \right)^2 dx}. \tag{4.1}$$

Now we recall that a classical measure of minimal position uncertainty is given by the "Fisher length" [11]

$$\delta x = \left[ \int P \left( \frac{\nabla P}{P} \right)^2 dx \right]^{-1/2}. \tag{4.2}$$

Comparing (4.1) and (4.2) immediately provides the "exact uncertainty relation" of reference [4],

$$\delta x = \frac{1}{\sqrt{\int P \left( \frac{\hbar}{2} \frac{\nabla P}{P} \right)^2 dx}} \frac{\hbar}{2} = \frac{1}{\delta p_0} \frac{\hbar}{2},$$

such that

$$\delta x \delta p_0 = \frac{\hbar}{2}. \tag{4.3}$$

This exact uncertainty relation holds only in a limiting case, however. In fact, if we now admit the general uncertainty in our knowledge of the total momentum, $\Delta p$, to come from both momentum contributions involved, i.e., according to equation (2.6),

$$\Delta p := \delta(\nabla S) + \delta p_0, \tag{4.4}$$

we obtain that

$$\Delta p \geq \delta p_0. \tag{4.5}$$

Moreover, according to the Cramer-Rao inequality of statistical estimation theory [12], it holds that the variance of any estimator $\Delta x$ is equal to, or larger, than the optimal variance, which is given by the Fisher length, i.e.,

$$\Delta x \geq \delta x. \qquad (4.6)$$

Therefore, combining equations (4.3), (4.5), and (4.6), one obtains Heisenberg's uncertainty relations

$$\Delta x \Delta p \geq \frac{\hbar}{2}. \qquad (4.7)$$

Thus, the uncertainty relations are physically explained by the "smearing out" of a particle's classical momentum due to the "osmotic" process of the zero-point field. Moreover, the form of equation (4.1) already hints at the recently established result [13] that the uncertainty relations are but a special consequence of the more powerful general statement that a quantum state is (nonlocally) entangled with the apparatus. Since the osmotic velocity in (4.1) depends only on the *relative gradient* of $P$, its expression does not necessarily fall off with any distance between component parts of a probability distribution. In other words, even small relative changes may become fully effective across nonlocal distances.

Concluding, one can envisage a new look on quantum mechanics by establishing a close link to the formalism of classical physics. In this way, also the essential differences to the latter can be elaborated thoroughly. In particular, as opposed to Nelson's attempts [2,3], the present approach also makes it possible to study in detail what it means that quantum mechanics is a theory with distinct nonlocal features. Finally, the approach presented here also leaves the options to consider more restrictive alternatives to the assumed extremal principle of equation (1.3), thus pointing towards a possible broader basis than the one which is strictly needed to obtain the orthodox quantum theory.